\documentclass[12pt,preprint]{aastex}
\shortauthors{Dikpati et al.}
\shorttitle{EnKF data assimilation in solar dynamo}

\begin{document}

\setkeys{Gin}{draft=false}

\title{Ensemble Kalman filter data assimilation in a 
Babcock-Leighton solar dynamo model: an observation system 
simulation experiment for reconstructing meridional flow-speed}

\author{MAUSUMI DIKPATI}
\affil{High Altitude Observatory, National Center for Atmospheric
Research \footnote{The National Center
for Atmospheric Research is sponsored by the
National Science Foundation. },
3080 Center Green, Boulder, Colorado 80301; dikpati@ucar.edu}
\author{JEFFREY L. ANDERSON}
\affil{IMAGe, National Center for Atmospheric
Research,
1850 Table Mesa Drive, Boulder, Colorado 80305}
\author{DHRUBADITYA MITRA}
\affil{NORDITA, Roslagstullsbacken 23 106 91 Stockholm, Sweden}

\begin{abstract}

Accurate knowledge of time-variation in meridional flow-speed and 
profile is crucial for estimating a solar cycle's features, which 
are ultimately responsible for causing space climate variations. 
However, no consensus has been reached yet about the Sun's meridional 
circulation pattern observations and theories. 
By implementing an Ensemble Kalman Filter (EnKF) data assimilation 
in a Babcock-Leighton solar dynamo model using Data Assimilation 
Research Testbed (DART) framework, we find that the best 
reconstruction of time-variation in meridional flow-speed can be 
obtained when ten or more observations are used with an updating 
time of 15 days and a $\le 10\%$ observational error. Increasing 
ensemble-size from 16 to 160 improves reconstruction. Comparison 
of reconstructed flow-speed with ``true-state'' reveals that EnKF 
data assimilation is very powerful for reconstructing meridional
flow-speeds and suggests that it can be implemented for 
reconstructing spatio-temporal patterns of meridional circulation.

\end{abstract}


\section{Introduction}

In recent years simulations of Babcock-Leighton type flux-transport
(hereafter BLFT) solar dynamos in both
2D and 3D \citep{ws91, dc99, md14} demonstrated the crucial role 
the Sun's global meridional circulation plays in determining solar cycle
properties. Time variations in speed and profile of meridional circulation 
have profound influence on solar cycle length and amplitude. The recent 
unusually long minimum between cycles 23 and 24 has been explained by 
implementing two plausible changes in meridional circulation, (i) by 
implementing the change from a two-celled profile in latitude in cycles 22 
to a one-celled profile in cycle 23 \citep{dgdu10}, and (ii) by performing 
a vast number of simulations by introducing a flow-speed change with 
time during the declining phase of each cycle \citep{nmm11}. Accurately 
knowing the speed and profile variations of the meridional circulation would 
greatly improve prediction of solar cycle features.

The meridional circulation has been observed in the photosphere and inside
the upper convection zone in the latitude range from the equator to 
$\sim 60^{\circ}$ in each hemisphere \citep{u10, ketal13, zhao13,
roth13, ksj14}. However, the speed, pattern and time variations
of the circulation at high latitudes and in the deeper convection zone 
are not known from observations yet. Theoretical models of 
meridional circulation \citep{m05, d14} provide some knowledge, but 
the flow patterns derived from model outputs vary from model to model, 
primarily because of our lack of knowledge of viscosity and density 
profiles and thermodynamics in the solar interior, which are 
essential ingredients in such models. As differential rotation does not
change much with time compared to meridional circulation, in this first
study we focus on time variation of meridional flow-speed, using a set-up
similar to that used previously \citep{cd00, nmm11}. Since the 
meridional circulation is a specified parameter in kinematic BLFT dynamos 
and the dynamo solutions depend sensitively on the spatio-temporal 
patterns of this circulation, we ask the question: can we infer the 
meridional circulation (in both space and time) from observations of 
the magnetic field? The purpose of this paper is to describe an Ensemble 
Kalman Filter (EnKF) data assimilation in a 2D BLFT solar dynamo model 
for reconstructing meridional flow-speed as a function of time for 
several solar cycles. A subsequent paper will investigate the reconstruction 
of spatio-temporal patterns of meridional circulation in the solar 
convection zone.

Data assimilation approaches have been in use for several
decades in atmospheric and oceanic models, but such approaches 
have been implemented in solar and geodynamo models only recently. 
\citet{jbt11} introduced a variational data assimilation system 
into an $\alpha$-$\Omega$ type solar dynamo model to reconstruct 
the $\alpha$-effect using synthetic data. Very recently 
\citet{scb13} applied a variational data assimilation method to 
estimate errors of the reconstructed system states in a stratified 
convection model. A detailed discussion of data assimilation in 
the context of the geodynamo can be found in \citet{fournier10}.  

In a sequential data assimilation framework, a set of dynamical
variables at a particular time defines a "model state", which
is the time-varying flow speed in the context of the present paper. 
Scalar functions of these state variables that 
can also be observed using certain instruments are called "observation 
variables", which are magnetic fields here. More detailed terminology
for identifying data assimilation components with solar physics
variables is given in \S2. In brief, the goal of sequential data 
assimilation is to constrain the model state at each time-step 
to obtain model-generated observation variables that are as 
close to the real observations as possible. The basic framework is 
based on statistical multidimensional regression analysis, a 
well-developed method that has been applied in atmospheric
and oceanic studies (see \citet{a01, awzh05, ac07} for details).
The EnKF sequential data assimilation framework also allows adding model 
parameters to the set of model states and estimating values of these
parameters that are most consistent with the observations. 

It is a common practice to perform an ``Observation System Simulation 
Experiment'' (OSSE) in order to validate and calibrate the 
assimilation framework for a particular model. An OSSE generates
synthetic observations from the same numerical model that is used
in the assimilation. In this case the numerical model is a simple 
BLFT dynamo model containing only a weak nonlinearity in the 
$\alpha$-quenching term; thus adding Gaussian noise to 
model-outputs for producing synthetic observations works well.
In a more realistic situation for a large system with highly 
nonlinear processes, such as in numerical weather prediction models, 
it may be necessary to use a non-Gaussian ensemble filter (see, e.g. 
\citet{a10}).

A few examples of predicting model parameters using sequential data 
assimilation techniques have been presented by \citet{mfc04} and 
\citet{cmf04} in the context of estimating neutral thermospheric 
composition, and most recently by \citet{mla13} for estimating 
thermospheric mass density. An EnKF data assimilation framework has 
recently been applied to a 3D, convection-driven geodynamo model 
for full state estimation using surface poloidal magnetic fields as 
observations \citep{fna13}. We implement EnKF sequential data 
assimilation to reconstruct time-variations in meridional flow-speed 
for several solar cycles, using poloidal and toroidal magnetic fields 
as observations. We note certain differences in our case compared 
to the cases described above, namely, unlike neutral thermospheric 
composition and thermospheric mass density, the meridional flow-speed 
is not governed by a deterministic equation. 

\section{Methodology and Assimilation setup}

In order to describe the EnKF data assimilation methodology, we 
first identify the data assimilation components with solar physics
variables. A physical BLFT dynamo model, which generates magnetic 
field data for a given the meridional flow-speed, is called the 
``Forward Operator''. The time-varying meridional flow-speed at a 
given time is the ``model state'' and will be estimated by the EnKF 
system. The EnKF requires a prediction model that generates a forecast 
of the meridional flow-speed at a later time given the value at the 
current time. Here, that prediction model simply adds a random 
draw from a Gaussian distribution to the current meridional 
flow-speed, because the BLFT dynamo model we are using here is
a kinematic dynamo model. In the future, the results can be further
refined by imposing additional physical conditions using a 
dynamical dynamo model. It is important not to confuse the 
prediction model with the BLFT dynamo model that acts as a 
forward operator for the data assimilation process.

Figure 1 schematically depicts the data assimilation framework 
considering an ``ensemble of three members'' of the prediction 
model state (meridional flow-speeds). Sequential assimilation 
can usually be described as a two-step procedure, a forecast 
followed by an analysis each time an observation is available. 
In Figure 1, ${v_1}^t, {v_2}^t, {v_3}^t$ near the label (a) 
denote three different realizations of initial flow-speeds, which
are input to the prediction model. We generate ``prior'' estimates 
of the model state (denoted by ${v_1}^{{t+\delta}^{\rm \,prior}}, 
{v_2}^{{t+\delta}^{\rm \,prior}}, {v_3}^{{t+\delta}^{\rm \,prior}}$ 
near the label (b)) by using the prediction model to advance the 
estimates of the flow-speed from the initial time ($t$) to the 
time ($t+\delta$) at which the first observations of magnetic 
field are available. In this case, the prediction model just adds 
a different draw from the Gaussian noise distribution to each initial 
ensemble estimate of the flow speed ($\Gamma$ along the solid green 
arrows in Figure 1). The resulting ensemble of flow speeds is 
referred to as a prior ensemble estimate. The central equation 
for estimating the prior state is a stochastic equation given by,
$${v_1}^{{t+\delta}^{\rm prior}}={v_1}^{t}+\Gamma_0 \times \gamma, 
\quad\eqno(1), $$ 
in which, $\gamma$ is a function for generating normalized Gaussian 
random numbers with unit amplitude and unit standard deviation, and 
the amplitude of the prediction model noise is governed by $\Gamma_0$. 
Thus the evolution of the system from label (a) to (b) through 
Equation (1), denoted by 'Eq1' in Figure 1, constitutes the first 
step of the two-step procedure in sequential assimilation.

\clearpage

\begin{figure}[hbt]
\epsscale{1.0}
\plotone{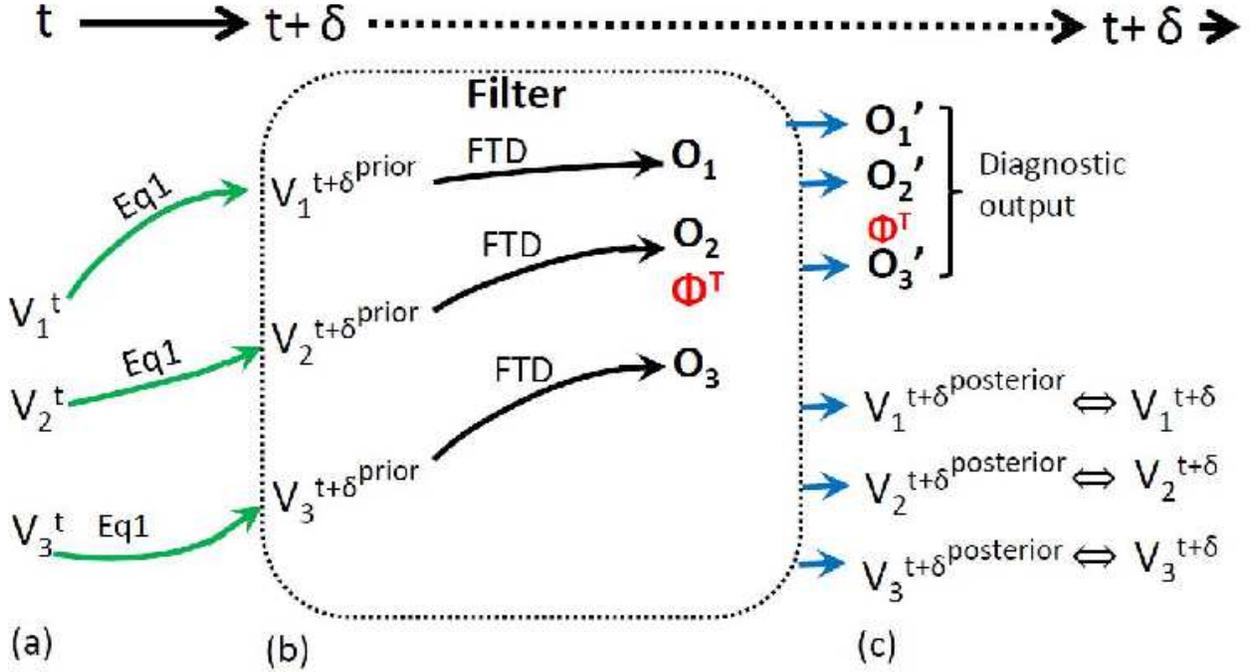}
\caption{
Schematic representation of the EnKF method as applied in
present problem. Three members of the ensemble of flow-speeds at
time $t$, namely $V_1^t$, $V_2^t$ and $V_3^t$ are evolved using
Equation (1) to generate corresponding prior state at time $t+\delta$,
shown as evolution from (a) to (b). In the ``Filter'' box, observations
$O_1$, $O_2$ and $O_3$ are generated by applying forward operator (FTD)
on the prior estimates. Then following a linear regression analysis
on these observations with corresponding flow-speeds, the ``Filter''
estimates posterior flow-speeds using the synthetic observation
with errors. In the process it also generates the posterior
observation for evaluation of innovation at time $t+\delta$.
Time advances from $t$ to $t+\delta$ (denoted by solid line at the top)
during the evolution from (a) to (b), while in the ``Filter'' procedure,
time remains frozen at step $t+\delta$ (denoted by dotted line from (b)
to (c)).
}
\label{Enkf-diagram}
\end{figure}

The second step includes (i) producing observations from outputs of the
forward operator and (ii) estimating posterior flow-speeds by employing 
regression among these observations, real observations (synthetic in
OSSE) and prior flow-speeds. The forward operator (BLFT dynamo in this 
case) is denoted by ${\rm FTD}$ along the solid black arrows, which uses
the prior estimates of flow speed to produce a prior ``ensemble of 
observation estimates'' which are magnetic field outputs. Three 
realizations of magnetic fields from the forward operator are denoted 
by $O_1$, $O_2$, $O_3$ in Figure 1. Note that the statement, ``forward 
operator operating on three model states generates three prior 
observation estimates'', is equivalent to the statement, ``BLFT dynamo 
running with three meridional flow speeds produces three sets of model 
outputs of magnetic fields''.  

To elucidate the second step of assimilation, we describe the function
of the ``Filter''. For real prediction using EnKF data assimilation, 
we would have real data (observations) from instruments; in our OSSE 
case it is synthetic, denoted by $\Phi^T$ in Figure 1. Synthetic 
observations are generated by applying the forward operator to a 
specified time series of meridional flow-speeds (as shown in Figure 2a). 
Our goal is to apply an EnKF to obtain an improved distribution of 
estimated flow-speeds (i.e. ``posterior states'', 
${v_1}^{{t+\delta}^{\rm \,posterior}}, {v_2}^{{t+\delta}^{\rm \,posterior}}, 
{v_3}^{{t+\delta}^{\rm \,posterior}} $) using the prior ensemble and the 
observation of magnetic field. The EnKF (black-dotted box in the 
diagram) first compares the prior ensemble of observation-estimates 
to the actual observation and computes increments to the prior 
observation-estimates. These observation-increments are then 
regressed using the joint prior-ensemble distribution of flow-speed
and magnetic field observations to compute increments for the 
prior-ensemble of flow-speeds. The EnKF can also produce a 
posterior-ensemble of magnetic field observations 
(${O_1}^{\prime}, {O_2}^{\prime}, {O_3}^{\prime} $), which can be 
used for diagnostic purposes. The posterior-ensemble distribution 
of flow-speeds is the best estimate of the flow-speed distribution 
at time $t+\delta$ given the available observations. Mean flow-speed 
at time $t+\delta$ can be calculated by taking the average over all 
ensemble members.

To proceed with the reconstruction of flow-speeds at the time of 
the next observation, $t+2\delta$, the reconstructed flow-speeds at 
time $t+\delta$ are used as the input to the prediction model 
(Equation (1)), and the same procedure described in the previous 
paragraphs is repeated. Random Gaussian noise through Equation (1)
prevents degeneration of ensemble. Thus after many time steps, the 
entire time series of the ensemble distribution of flow-speeds 
can be constructed. Time series of the mean flow-speeds can be 
calculated by taking the average over all ensemble members at 
each time. However, it may produce a better reconstruction in some 
cases if one ensemble member, which produced observation ($O^{\prime}$) 
closest to real observation ($\Phi^T$), is chosen.

In order to perform an OSSE, it is now necessary to define the
``true state'' flow-speed as a function of time. Synthetic 
observations are generated at selected times by applying the
forward operator (BLFT dynamo model) to the true state flow-speed
at the appropriate time and adding on a random draw from a specified
observational error distribution to simulate instrumental and 
other errors. 

\clearpage

\begin{figure}[hbt]
\epsscale{0.9}
\plotone{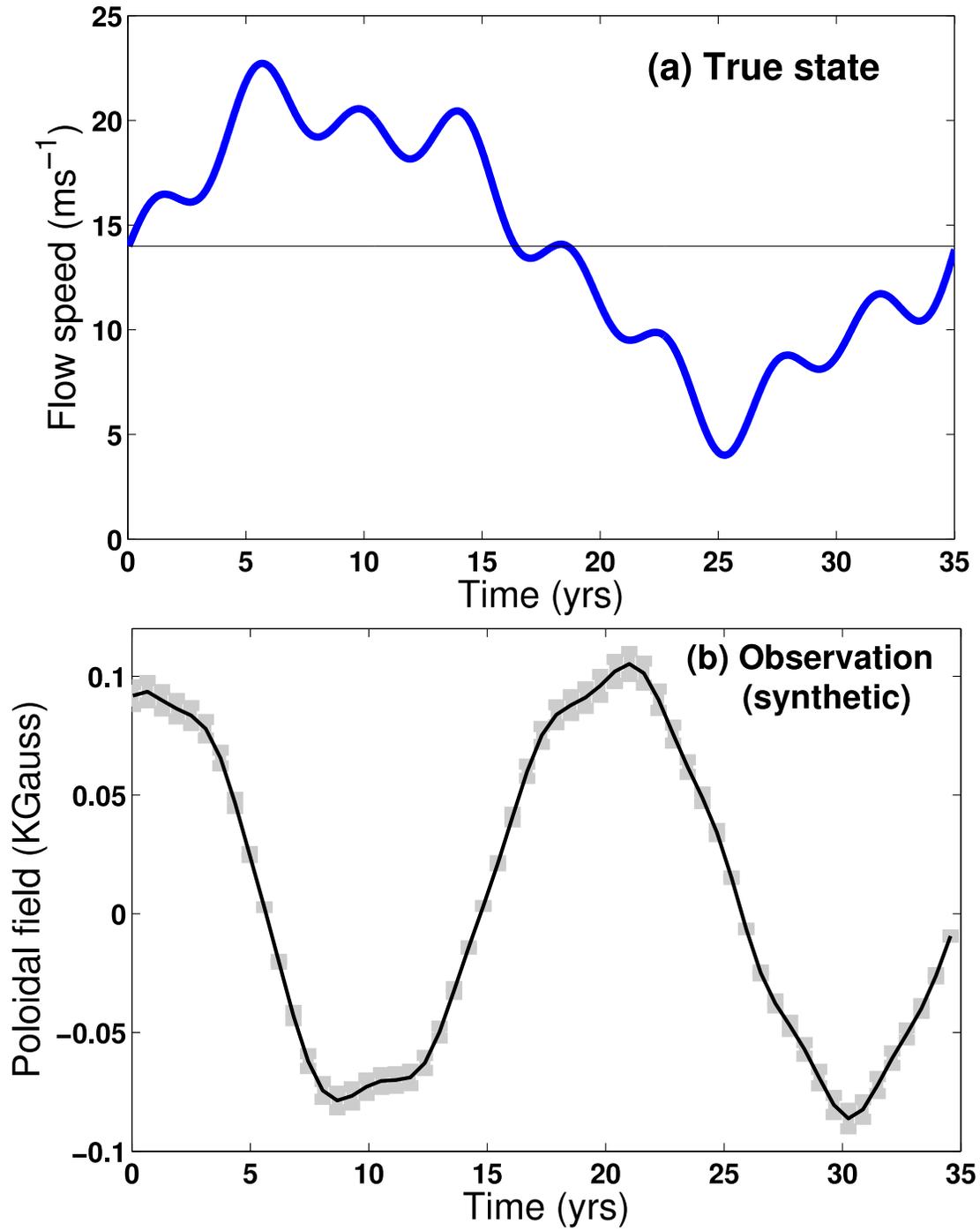}
\caption{Panel (a): true state (i.e. time-varying meridional flow speed)
that varies by $\pm70\%$ about mean flow-speed, $14\,{\rm m}{\rm s}^{-1}$;
Panel (b): synthetic observations with observational error bars, shown 
at selected finite intervals for clarity.
}
\label{truth-obs}
\end{figure}

\clearpage

As noted above, the forward operator is a kinematic
BLFT dynamo model, described in detail in \citet{dgdu10}. 
The dynamo ingredients are a solar-like differential rotation, 
a single-celled meridional circulation, a Babcock-Leighton type 
surface $\alpha$-effect and a depth-dependent diffusivity; the
mathematical forms are prescribed in \citet{dgdu10}. Dynamo 
equations, computation domain and boundary conditions are used 
as given in \citet{dgdu10}. At $t=0$ we start the integration
of the dynamo-DART system over the first assimilation-step
by initializing the forward operator (FTD model) with a 
converged solution for a flow-speed of $14\,{\rm m}{\rm s}^{-1}$. 
In the subsequent assimilation-steps, the solution at the end
of previous assimilation-step for each ensemble-member is used
as initial condition. 

We construct a time-varying flow-speed for a span of 35 years
(see Figure 2(a)), guided by observations \citep{ba10, u10}, 
which has a natural variation of 20-40\% with respect to the 
mean flow considered here, i.e. 14 ${\rm m}{\rm s}^{-1}$, 
as shown by the thin black line in Figure 2(a). This specified 
time-varying flow-speed is referred to as the true model state.
Thus, keeping the spatial pattern of meridional circulation fixed, 
we consider here reconstructing the time-series of the scalar 
flow-speed. 

To generate synthetic magnetic field data, we incorporate the 
time-varying true meridional flow-speed in our BLFT dynamo, and 
simulate the time series of idealized magnetic fields in the entire 
computation domain. Then we construct the time series of synthetic
magnetic observations by adding synthetic observational error to 
the simulated idealized magnetic data. Figure 2(b) shows a single 
observation, which is created from the simulated poloidal field by
extracting from the location, $r=0.98R$ and $\theta=86.5^{\circ}$. 
However, note that the dynamo simulation in $101\times101$ grids can 
give us as many as 20202 synthetic magnetic observations of poloidal 
and toroidal fields.

\section{Results}

Considering only one observation, as shown in Figure 2(b), we perform 
assimilation runs with 16 ensemble members with an observational 
error of $\sim 30\%$, which means an error of $\sim 30\%$
about the ideal magnetic field generated using the true meridional
flow speed. To estimate the prior states of flow-speed, we use
Equation (1), in which we set $\Gamma_0=0.5 {\rm m}{\rm s}^{-1}$.
If the meridional flow-speed varies up to $\pm 20\%$ (i.e.
$\sim \pm 2.8 {\rm m}{\rm s}^{-1}$ for a mean flow-speed of 
$14 {\rm m}{\rm s}^{-1}$) during six months, the variation in
15 days can be $\sim 0.23 {\rm m}{\rm s}^{-1}$. Thus we chose
$\Gamma_0=0.5 {\rm m}{\rm s}^{-1}$ so that it is large enough to 
capture the variation in flow-speed within our selected updating 
time-step of 15 days and also large enough to avoid ensemble collapse, 
but not so large as to produce unusual departures from cyclic behavior 
in a flux-transport dynamo. We show in Figure 3 the reconstructed 
meridional flow-speed as a function of time (panel (a)) and the 
estimates of the observation computed from the flow speed ensemble 
after data assimilation (panel (b)). 

We see in Figure 3(a) that the reconstruction is reasonably good 
except for the time windows between $\sim 15$ to 18 years and 33 to 
35 years during which we find $\sim 40\%$ error in the reconstructed 
flow speed. Observations indicate $\sim 10 - 20 \%$ error in the 
measurement of meridional flow speed \citep{ba10, u10}. The inset of 
Figure 3(a) reveals, for an initial guess, far-off from the true-state, 
the reconstructed states asymptotically converges towards the true-state. 
Even though they oscillate around the truth with large amplitude,
the oscillations damp with time. Figure 3(b) shows histograms for the 
normalized distribution of flow-speeds before (in cyan) and after (in 
magenta) the analysis stage, along with true-state (blue), for the 
time instances of 5, 6.9, 10.1 and 27.5 years (marked by vertical 
lines in Figure 3(a)) during assimilation. For this case with 16 
ensemble members we chose 10 bins for an optimal display. Four 
time instances are chosen in such a way as to present the 
following four different phases of reconstructions: (i) the 
distribution of prior and posterior states has small overlap (top 
left frame of Figure 3(b)), (ii) distribution is sharply peaked 
in one bin each for prior and posterior (top right frame), (iii) 
distribution is broad and has significant overlap (bottom left
frame) and (iv) distribution has no overlap at all (bottom right 
frame). However, we can clearly see the successful performance 
of the EnKF which reveals that in all cases the analysis phase 
brings the posterior distribution (magenta bars) closer to 
true-state (blue).

Figure 3(c) reveals that the assimilated magnetic observation (blue 
solid curve) is well-reproduced when compared with real observation 
(red-dashed curve). This is not surprising, because it has already 
been noted that short-term small fluctuations in flow speed do not 
significantly influence the overall evolution of global magnetic 
fields generated by a Babcock-Leighton dynamo (see, e.g. \citet{bd13}). 
We also plot the innovation ($I_i$) for this observation (black curve) 
and the cumulative innovation (orange curve). The innovation at the
$i^{th}$ analysis-step ($I_i$) is defined as the signed difference 
between real and reconstructed observations, whereas the cumulative 
innovation ($CI_i$) at $i^{th}$ analysis-step is the normalized sum of 
norm of innovation vectors over all the previous analysis-steps. $I_i$ 
and $CI_i$ being small, both have been ten-fold magnified to superimpose 
on observations. We see that, at three different time instances 
($t\approx 0$, 17 and 27 years), the innovation is relatively large; 
this is because at $t=0$ the initial guess is far-off from the truth, 
and at $t\approx 17$ and 27 years there are relatively sharp changes 
in flow-speed. The cumulative innovation asymptotes to zero as expected,
implying no bias of the system. To investigate the possibility of further 
improvement in the quality of the EnKF reconstruction, we examine the 
consequences of three important aspects of the EnKF: variation in 
observational error, size of ensemble and number of observations.  

\clearpage

\begin{figure}[hbt]
\epsscale{0.45}
\plotone{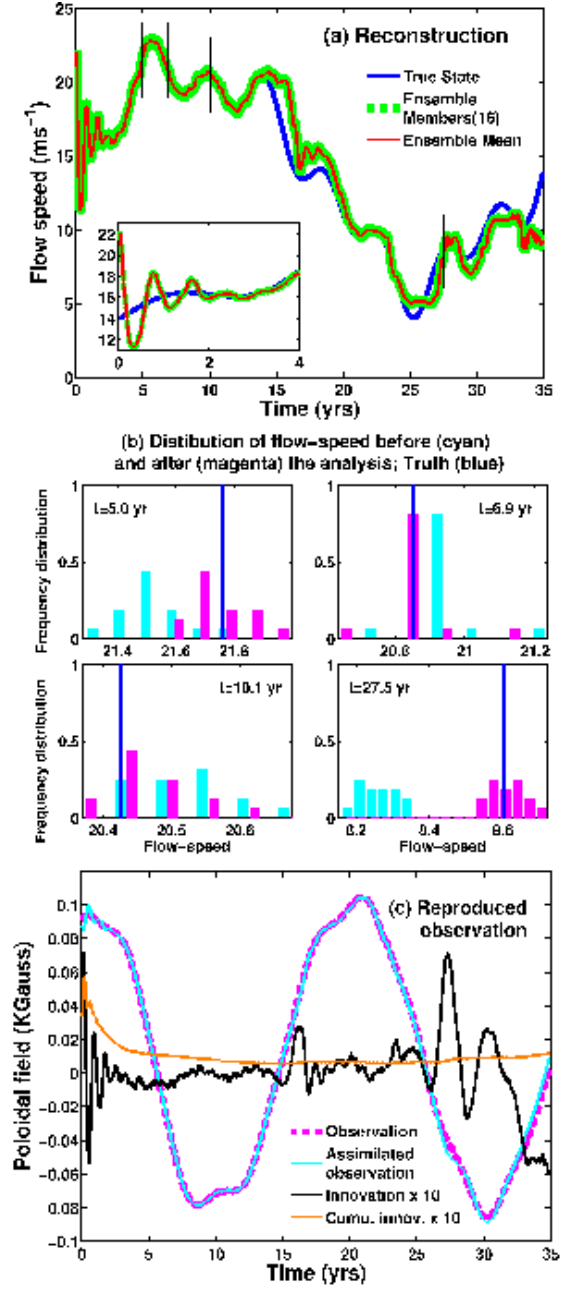}
\caption{In panel (a), red curve denotes ensemble-mean of
reconstructed meridional flow-speeds as a function of time; 
superimposed green curves denote all 16 ensemble members; true state 
is plotted in blue for comparison. Panel (b) shows normalized 
distribution of flow-speed in a histogram plot for four different 
times. The prior, posterior and true state are plotted in cyan, 
magenta and blue respectively. Panel (c) shows estimated observation 
after assimilation (i.e. poloidal magnetic field at $r=0.98R$ 
and $\theta=86.5^{\circ}$) in blue and actual observation 
in red; superimposed on that are innovation (black) and cumulative 
innovation (orange) (both ten-fold magnified for clarity). 
}
\label{reconstruction1}
\end{figure}

\clearpage

\begin{figure}[hbt]
\epsscale{0.5}
\plotone{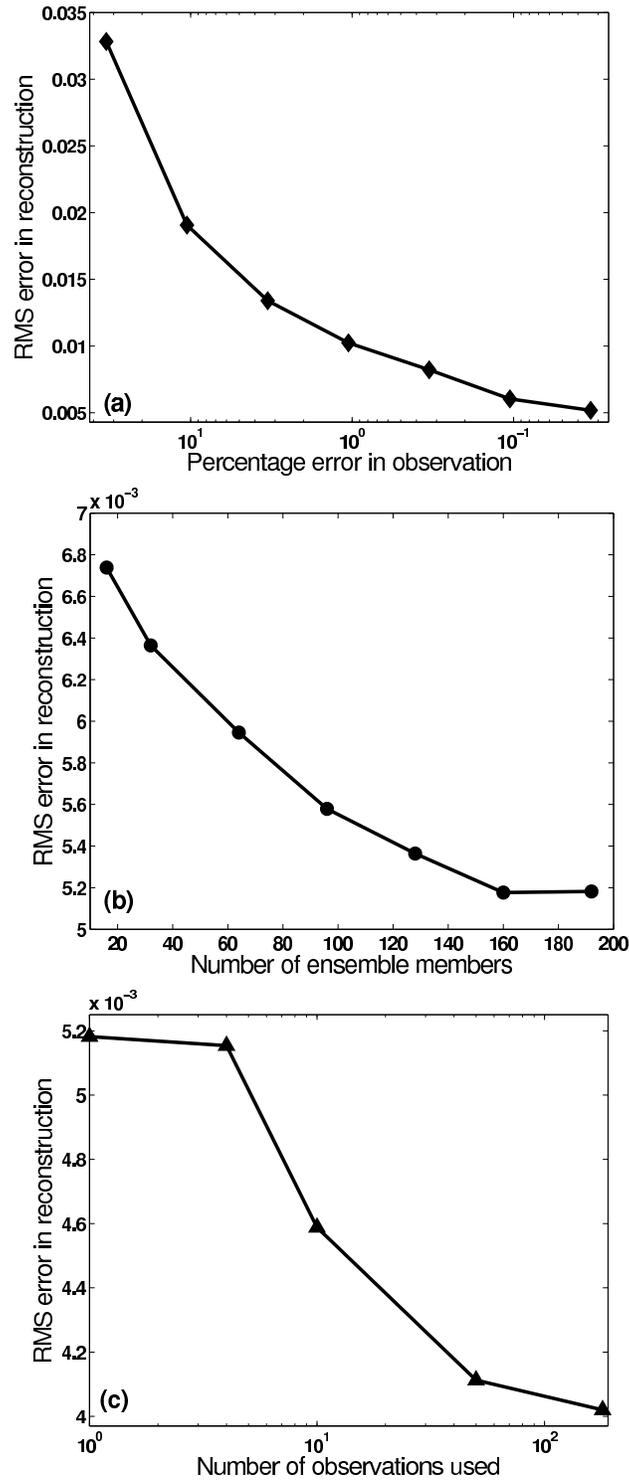}
\caption{Panels (a, b, c) show errors in reconstruction (in units of
${\rm m}{\rm s}^{-1}$) respectively as functions of observational errors 
(a), size of ensemble (b) and number of observations (c). 
}
\label{error-estimates}
\end{figure}
\clearpage

We perform convergence tests by estimating the error in the reconstructed
state as functions of observational errors, size of ensemble and 
number of observations. We define the error as the root mean square 
of the difference between the reconstructed state and the true-state. 
The assimilation interval is chosen to be 15 days in the 
present case; denoting every 15 days by the index $i$, the true 
state and the reconstructed state at the $i^{th}$ assimilation step 
by ${V_T}^i$ and ${V_R}^i$ respectively, we define the error as,
$ \sum_{i=1}^{n_{\rm max}}\sqrt{{({V_T}^i - {V_R}^i)}^2}/n_{\rm max} $,
in which $n_{\rm max}$ is the total number of indices for the
15-days assimilation intervals during the entire time-span of 35 years.
 
Figure 4 shows the rms errors in reconstructed flow speed as functions
of observational error (Figure 4(a)), size of ensemble (Figure 4(b)) 
and number of observations (Figure 4(c)). We see in Figure 4 that the 
error decreases systematically and asymptotes for certain values of the 
observational errors (1\%), size of ensemble (192 members) and number of 
observations (180). In the case of more than one observation, we
include more poloidal field observations (synthetic) from various
locations at and near the surface, and more toroidal field observations
at and near the bottom of the convection zone. While we vary observational
errors in panel (a), we use $1\%$ observational error in panels
(b) and (c). Panels (a,b) show convergence to typical hyperbolic 
patterns, as is often seen in numerical convergence tests (see, e.g. 
figure 3 of Dikpati (2012)). But panel (c) shows that the reconstruction 
can be improved systematically only when there are more than a certain 
number of observations (four in our case). In all these experiments, we 
used the same EnKF scheme. 

Bias or systematic error in an OSSE reconstruction may arise primarily 
because of the following assumptions made in the assimilation system: 
(i) the evolution of the ensemble spread is linear, (ii) the ensemble is 
sufficiently large, (iii) the forward operators are linear. Though these 
assumptions are roughly valid, they are not strictly true. But in general,
the resulting systematic errors are small unless the assimilation is 
applied in a large, highly nonlinear system like a numerical weather 
prediction model.
  
\begin{figure}[hbt]
\epsscale{0.75}
\plotone{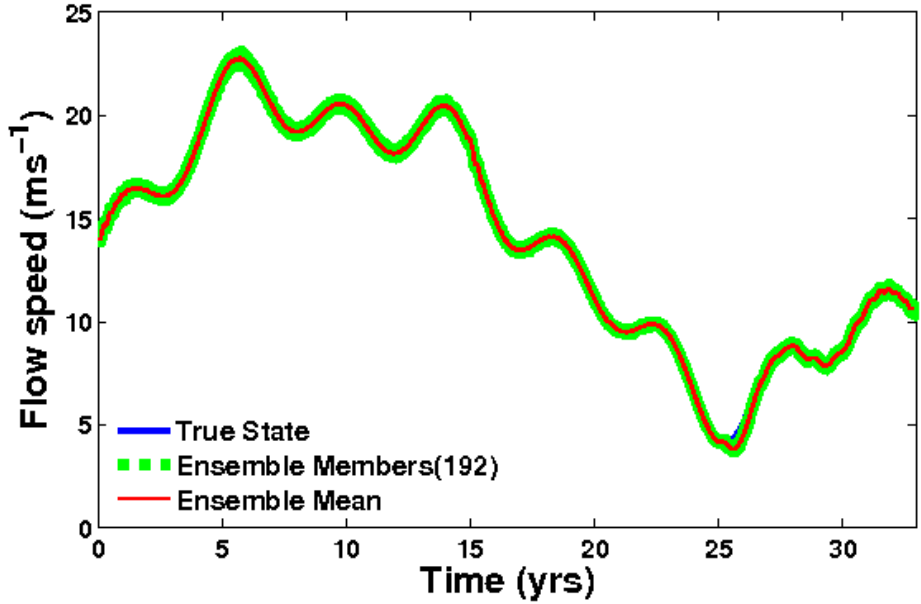}
\caption{As in Figure 3(a), red and green curves denote reconstructed 
meridional flow-speed, 192 reconstructions for 192 ensemble members
as function of time. Total 180 observations with observational error
of $1\%$ have been used. Note that the true state (blue) is invisible, 
hidden behind the red and green curves.
}
\label{reconstruct-final}
\end{figure}

With the knowledge gained from the convergence experiments shown in 
Figure 4, we consider a case with 192 ensemble members and 180 magnetic
observations, consisting of equal numbers of poloidal and toroidal
magnetic fields at various latitude and depth locations. Each of these 
observations has an observational error of $1\%$. We present the 
reconstructed flow-speed from this assimilation in Figure 5. The 
reconstructed flow-speed (red curve) matches very well with the true 
state, and thus the true state plotted in blue is essentially hidden 
behind red and green curves. It is not realistic to expect an 
observational error of as small as 1\%; Figure 5 presents here
an illustrative example of one of the best possible reconstructions.
In fact, several additional assimilation runs indicate that the
reconstruction is still good if the observational error does not 
exceed 40\%, and reasonably good when 90 out of 180 observations
have up to $\sim 50\%$ errors. But the reconstruction fails when all
observations have more than 40\% error.

What we have demonstrated so far is that the time-dependent amplitude 
of meridional circulation, having one flow-cell per hemisphere, can be 
reconstructed by implementing EnKF data assimilation with synthetic
data. In reality we do not know from observational data whether they 
were produced by a dynamo operating with a single-celled flow in each 
hemisphere, or with a more complex flow profile, or with a combination
of complex time-variations in all possible dynamo ingredients. In order
to investigate the outcome from this method when the assumption made
about the flow profile is wrong, we carry out an experiment to 
reconstruct flow-speed assuming a one-cell flow, while using observations
of magnetic fields produced from a flow pattern that has two cells in
latitude in each hemisphere (see \citet{dgdu10} for prescription of a
two-celled flow). We obtain synthetic data for a case with two flow 
cells in latitude. Using 192 ensemble members and 180 magnetic 
observations with 1\% error in each of these observations, we 
estimate the time variation in flow-speed by assuming 
a single-celled flow, and plot in Figure 6(a). We find that the 
reconstruction is relatively poor, as expected. However, with a closer 
look we can see that the reconstructed speed is trying to approach 
the true-state from a lower value for the first 12 years and from a 
higher value for the next 13 years. When the trend in the true-state
reverses near the year 27, the OSSE has greater difficulty in converging
on it. 

\begin{figure}[hbt]
\epsscale{0.58}
\plotone{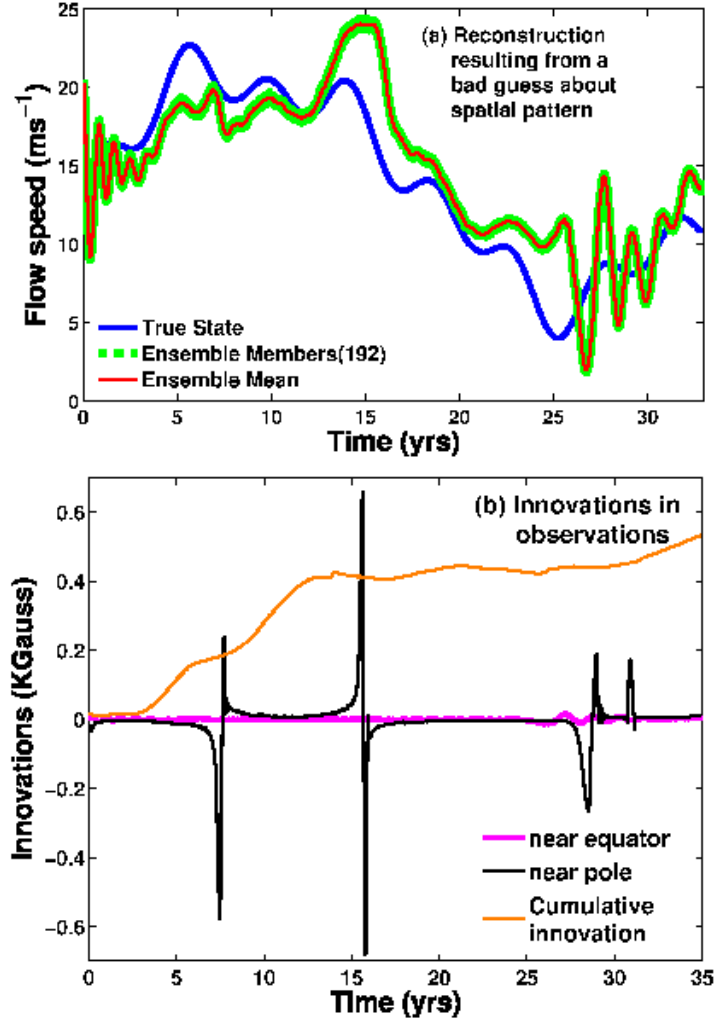}
\caption{(a) Same as in Figure 5, but for a case in which assumption
made about spatial structure of flow is incorrect. (b) Two characteristic
innovations for observations at $\theta=86.5^{\circ}$ and $3.6^{\circ}$
are plotted respectively in magenta and black curves, and cumulative
innovation in orange curve.  
}
\label{case-incorrect-assumption}
\end{figure}

Figure 6(b) shows the innovation for two typical observations of 
poloidal fields near the surface, at $\theta=86.5^{\circ}$ (i.e.
at low latitude near the equator) and at $3.6^{\circ}$ (at high
latitude near the pole). Much larger innovation at high latitude
than at low latitude reflects the fact that the spatial pattern 
is more erroneous at polar latitudes. The cummulative innovation
(orange curve) does no longer asymptote to zero; this implies
bias in the system, due to incorrect assumption of flow-pattern.
In the future, we will extensively explore the reconstruction of 
spatial pattern of meridional circulation.  

\section{Comments and conclusions}

We have demonstrated through OSSEs that EnKF data assimilation
into a Babcock-Leighton flux-transport dynamo can successfully 
reconstruct time-varying meridional circulation speed for several 
cycles from observations of the magnetic field. To obtain the best 
reconstruction, we have fed the assimilation system of 192 ensemble 
members with 180 observations with 1\% observational error. However,
a reasonably good reconstruction can be obtained when all observations
have up to $40\%$ error, or half of the observations have up to $50\%$ 
errors, but the rest of them have much smaller errors.

\citet{da12} noted that the response time of a Babcock-Leighton 
dynamo model to changes in meridional flow is $\sim 6$ months, so
the relatively poor reconstruction with only one observation with 
33\% observational error and 16 ensemble members can be improved if 
this information about the dynamo model's response time to flow changes 
can be exploited. A forthcoming paper will investigate how
to use this response time during assimilation. 

Throughout this paper, we used an assimilation interval of 15-days. 
The reason for this choice is as follows.
Recently \citet{sfa14} have done a very thorough assessment of the
'predictability' of solar flux-transport dynamos. Predictability of a
model refers to the time it takes for two solutions that start from
slightly different values of either initial conditions or input
parameters to diverge from each other to the point that they forecast
substantially different outcomes. \citet{sfa14} found an $e$-folding
time of about 30 years. The flux-transport dynamo model we use
here is physically very similar to theirs. Therefore we judge that 
the $e$-folding time for our model will be similar.

It is clear that the time interval for updating the data in our OSSE's
should be much less than the predictability limit of the model, but it
should be long compared to the integration time step (a few hours), and
consistent with the time scale for changes in axisymmetric solar 
observations, which is one solar rotation. It should also be shorter 
than the "response time" of our dynamo model ($\sim 6$ months, see 
\citet{da12}) to a sudden change in inputs. We have therefore chosen 
our updating interval to be 15 days; we plan to test the sensitivity 
of assimilation to changes in the updating interval from 15 days to 
longer (for instance, 30 days) and shorter ($\sim 3$ days). 

In this study we have demonstrated what it takes to reconstruct the 
amplitude variations with time of a one-celled meridional circulation
of fixed profile with latitude and radius. In reality the Sun's 
meridional circulation may not be a one-celled pattern -- it may 
undergo changes both in profile and speed with time. We have
demonstrated that the innovation in observation-forecast can be
very large when the assumption about the spatial structure of the
flow-pattern is incorrect, and it can be even larger where
the departure in assumed spatial pattern from actual pattern in 
flow is larger. Thus an obvious next step with our data assimilation 
system would be to attempt to reconstruct the spatio-temporal 
pattern of meridional flow.

Our ultimate goal is to perform assimilation runs from actual 
observations instead of synthetic data, in cases of reconstruction 
as well as future predictions. From this study we can build confidence 
about the power of EnKF data assimilation for reconstructing not only 
the flow speed but also the profile of meridional circulation in 
the entire convection zone of the Sun in the future.

\begin{acknowledgements}
We thank Nancy Collins and Tim Hoar for their invaluable help with 
assimilation tools and software. We extend our thanks to two 
reviewers for many helpful comments and constructive suggestions, 
which helped significantly improve the paper. The DART/Dynamo 
assimilation runs have been performed on the Yellowstone Supercomputer 
of NWSC/NCAR under project number P22104000, and all assimilation tools 
used in this work are available to the public from 
$http://www.image.ucar.edu/DAReS/DART$. This work is partially supported 
by NASA grant NNX08AQ34G. Dhrubaditya Mitra was supported by the European
Research Council under the AstroDyn Research Project No. 227952 and the 
Swedish Research Council under grant 2011-542 and the HAO visitor program. 
National Center for Atmospheric Research is sponsored by the National 
Science Foundation. 
\end{acknowledgements}


\end{document}